\documentstyle[12pt]{article}
\newcommand{\prepr}[1] {\begin{flushright} {\bf #1} \end{flushright} \vskip
1.5cm}
\newcommand{\titul}[1] {\begin{center}{\large\bf #1 } \end{center}\vskip 1.cm}
\newcommand{\autor}[1] {\begin {center} {\large \lineskip .5em #1 }
                        \end   {center} }
\newcommand{\lugar}[1] {\begin{center} {\it #1} \end{center}}
\newcommand{\abstr}[1] {{\begin{center} \vskip .5cm {\bf Abstract
                        \vspace{0pt}} \end{center}}\begin{quote} #1
                        \end{quote}}
\newcommand{\z}{&&\hspace*{-1cm}}
\begin{document}

\begin{titlepage}
\prepr{ENSLAPP-A-573/95\\ US-FT-29-95\\ December 1995}
\titul{The Gluon Distribution as \\a Function of F$_2$ and dF$_2$/dlnQ$^2$
at small x.\\ The Next-to-Leading Analysis}
\autor{A.V. Kotikov\footnote{E-mail:KOTIKOV@LAPPHP8.IN2P3.FR}}
\lugar{Laboratoire de Physique Theorique ENSLAPP\\ LAPP, B.P. 100,
F-74941, Annecy-le-Vieux Cedex, France}
\autor{G. Parente\footnote{E-mail:GONZALO@GAES.USC.ES}}
\lugar{Departamento de F\'\i sica de Part\'\i culas\\
Universidade de Santiago de Compostela\\
15706 Santiago de Compostela, Spain}
\abstr{We present a set of formulae to extract the gluon
distribution function from the deep inelastic structure function F$_2$ and
its derivative dF$_2$/dlnQ$^2$ at small x in the leading
and next-to-leading order of perturbation theory. The detailed analysis is
given for new HERA data. The values of the gluon distribution are found
in the range $10^{-4} \leq x \leq  10^{-2}$ at $Q^2=20$ GeV$^2$}
\end{titlepage}
\newpage

The knowledge of the DIS structure functions at small values
of the Bjorken scaling variable $x$ is interesting for
understanding
the inner structure of hadrons. Of great relevance is the
determination of the gluon density at low x, where gluons
are expected
to be dominant, because it could be a test of perturbative QCD
or a probe of new effects,
and also because it is the basic ingredient in
many other calculations of different high energy hadronic
processes.

Recently two experiments working in the electron-proton
collider HERA at DESY (H1 and ZEUS) have published
new data on the structure function $F_2$ \cite{F2H1}, \cite{F2ZEUS}.
Up to now all the analysis performed of these data
\cite{H1PREP95} -
\cite{MRS95} found
that the gluon distribution rises
steeply towards low x (in the moderate Q$^2$
range of the measurements). This behaviour has been recently connected
within the DGLAP evolution equations \cite{KOTILOWX}
with the less singular are found at lower values of Q$^2$ by NMC and E665
experiments.

  We introduce the standard parameterizations of singlet quark
$s(x,Q^2)$ and gluon $g(x,Q^2)$ parton distribution functions
(PDF)\footnote{We use PDF multiplied by $x$
($s=xS$ and $g=xG$)
 and neglect the nonsinglet quark distribution at small $x$}
  (see, for example, \cite{MRS95}) \begin{eqnarray} s(x,Q^2) & = & A_s
x^{-\delta} (1-x)^{\nu_s} (1+\epsilon_s \sqrt{x} + \gamma_s x)
 \equiv x^{-\delta} \tilde s (x,Q^2)  \nonumber \\ g(x,Q^2) & = & A_g
x^{-\delta} (1-x)^{\nu_g} (1+\epsilon_g \sqrt{x} +\gamma_g x) \equiv
x^{-\delta} \tilde
 g(x,Q^2) ,\label{1} \end{eqnarray}
with $Q^2$ dependent parameters in the r.h.s..

Note that the behaviour of Eq. (\ref{1}) with a $Q^2$-independent value for
$\delta ~ (\delta_q = \delta_g)$ obeys the DGLAP equation
when $x^{-\delta} \gg 1$ (see,  for example, \cite{Mar} -
\cite{EKL}). If $\delta (Q^2_0) =0$
in some point $Q^2_0 \sim 1 GeV^2$ (see \cite{F2NMC}, \cite{BF},
\cite{KOTILOWX}) , then the behaviour $p(x,Q^2) \sim Const$
($p=(s,g)$) is not compatible with DGLAP equation and a more singular
behaviour is generated. If we restrict the analysis to a Regge-like
form of structure functions, one obtains (see \cite{KOTILOWX})
$$ p(x,Q^2) \sim x^{-\delta_p(Q^2)} $$
with next-to-leading order (NLO) $\delta_q(Q^2) \neq
\delta_g(Q^2)$ intercept trajectories.

Without any restriction the double-logarithmical behaviour, i.e.
\begin{eqnarray}
 p(x,Q^2) \sim \exp{(\frac{1}{2} \sqrt{\delta_p(Q^2)ln\frac{1}{x}})}
\label{2} \end{eqnarray}
is generated. At NLO and for $f=3$ active quarks,
$\delta_p(Q^2)$ have the form (see \cite{KOTILOWX}):
$$\delta_g(Q^2)~=~ \frac{4}{3} t -\frac{1180}{81} l,~~
\delta_q(Q^2)~=~\delta_g(Q^2)  -20 l$$
while for $f=4$ one has:
$$\delta_g(Q^2)~=~ \frac{36}{25} t -\frac{91096}{5625} l,~~
\delta_q(Q^2)~=~\delta_g(Q^2)  -20 l$$
where
$t~=~ln(\alpha(Q^2_0)/\alpha(Q^2))$ and $l~=~ \alpha(Q^2_0) - \alpha(Q^2)$.

Because we would like to get formulae to extract the gluon distribution
from experimental data without theoretical restrictions, we will consider
both, the Regge-like behaviour of Eq. (\ref{1}) if $x^{-\delta} \gg 1$,
and the non-Regge-like behaviour of Eq. (\ref{2}) if $\delta(Q^2_0)=0$.

In this work we present a simple relation between the gluon and
F$_2$ which permits the extraction of the gluon directly from data
with a NLO precision in perturbative QCD.
This kind of formulae (see also the NLO result of ref. \cite{PRYTZ}
which depends on an unknown phenomenological function) are useful
as an alternative to the most complex analysis involved in global QCD fits.
At low x the coupled integro-differential equations
can be converted in  more simple linear relations between
parton densities and structure functions\footnote{There
are analogous formulae connecting  $F_L$ with $F_2$
and its derivative
and also for the extraction of
gluons from $F_L$ (see the at LO
refs.\cite{KOTIJETP95} and \cite{COOPER} and at NLO
refs. \cite{INPROGRESS} and \cite{KOTIPRD94}, respectively)}.
The method to arrive to the solution presented below is based in the
replacement of the Mellin convolution by ordinary products developed
in ref. \cite{KOTIPRD94} that was already applied in the derivation
of the leading order (LO) formula in ref. \cite{KOTIJETP94} and the
NLO result at
$\delta =1/2$ in ref. \cite{KOPA}.\\

{\bf 1.}
 Assuming the {\it Regge-like behaviour} for the gluon distribution and
 $F_2(x,Q^2)$:
$$g(x,Q^2)  =  x^{-\delta} \tilde g(x,Q^2),~~
F_2(x,Q^2)  =  x^{-\delta} \tilde f(x,Q^2), $$
 we obtain the
following equation for the $Q^2$ derivative of
the SF $F_2$\footnote{Hereafter contrary to the standard case we
use $ \alpha(Q^2)= \alpha_s(Q^2)/{4 \pi}$ .}:
 \begin{eqnarray} \frac{dF_2(x,Q^2)}{dlnQ^2}  =
-\frac{1}{2}
x^{-\delta} \sum_{p=s,g}
\Bigl(
 r^{1+\delta}_{sp}(\alpha) ~\tilde p(0,Q^2) +
 r^{\delta}_{sp}(\alpha)~ x \tilde p'(0,Q^2)  +
 O(x^{2}) \Bigr) ,\label{2.1} \end{eqnarray}
 where $
 r^{\eta}_{sp}(\alpha) $ are the combinations
of the anomalous dimensions (AD) of Wilson operators
$\gamma^{\eta}_{sp}= \alpha \gamma^{(0),\eta}_{sp} + \alpha ^2
\gamma^{(1),\eta}_{sp} + O(\alpha ^3)$
and Wilson
coefficients\footnote{Because
  we consider here $F_2(x,Q^2)$ but not the singlet quark
  distribution} $ \alpha B_2^{p,\eta}  + O(\alpha ^2)$
of the $\eta$
"moment"  (i.e., the corresponding variables expanded
from integer values of argument to non-integer ones):
 \begin{eqnarray} \z
r^{ \eta}_{ss}(\alpha) ~=~ \alpha
  \gamma^{(0),\eta}_{ss} + \alpha^2 \biggl(
  \gamma^{(1),\eta}_{ss} + B_2^{g,\eta} \gamma^{(0),\eta}_{gs} +
  2\beta_0 B_2^{s,\eta}
  \biggr),  \label{2.2}  \\  \z
r^{\eta}_{sg}(\alpha) ~=~ \frac{e}{f} \biggl[ \alpha
  \gamma^{(0),\eta}_{sg} + \alpha^2 \biggl(
  \gamma^{(1),\eta}_{sg} + B_2^{g,\eta} \bigl(2\beta_0 +
  \gamma^{(0),\eta}_{gg} - \gamma^{(0),\eta}_{ss} \bigr) \biggr)
  \biggl] \nonumber  \end{eqnarray}
 and
$$ \tilde p'(0,Q^2) \equiv \frac{d}{dx} \tilde p(x,Q^2) \mbox{ at }
x=0$$
where $e = \sum_i^f e^2_i$ is
the sum of squares of
quark charges.

For the gluon part from r.h.s of Eq.(\ref{2.1}) with accuracy of $O(x^2)$,
we have the form:
 \begin{eqnarray}
r^{1+\delta}_{sg} \tilde
   g(x/ \xi_{sg},Q^2)~
 \mbox{ with } ~ \xi_{sg} =
r^{1+\delta}_{sg} /r^{\delta}_{sg}
 \label{3} \end{eqnarray}

In the quark part a similar simple form is absent because the
corresponding LO anomalous dimensions, $\gamma^{(0),1+\delta}_{ss}$ and
$ \gamma^{(0), \delta}_{ss} $, have opposite signs. However, within
accuracy
$O(x^2)$, it may be represented as a sum of two terms like Eq.(\ref{3}),
with a shift of some coefficients and arguments. Choosing the shifts as 1 and
A we have the following representation for the quark part:
$$c_1~\tilde s(x,Q^2) ~ +~ c_2~\tilde s(Ax,Q^2) ~+~ O(Ax^2), $$
where
 \begin{eqnarray}
 c_2 ~=~ \frac{r^{\delta}_{ss}-r^{1+\delta}_{ss}}{A-1}
{}~ \mbox{ and } ~c_1 ~=~ \frac{A r^{1+\delta}_{ss}-r^{\delta}_{ss}}{A-1}
 \label{4} \end{eqnarray}

We thus find  the following expression from Eqs.
(\ref{2.1})-(\ref{4})
 \begin{eqnarray}    \z    \frac{dF_2(x,Q^2)}{dlnQ^2}  = -\frac{1}{2} \biggl[
r^{1+\delta}_{sg} {(\xi_{sg})}^{-\delta}
   g(x/\xi_{sg},Q^2) +
c_1F_2(x,Q^2)  + c_2F_2(Ax,Q^2) \biggr] \nonumber \\ \z
{}~~~~~~~ ~~~~~~~~
+~ O(x^{2-\delta}, \alpha A x^{2-\delta}) ,\label{5} \end{eqnarray}
because $c_i \sim O(\alpha)$.

{}From Eq. (\ref{5}) with the accuracy of $O(x^{2-\delta}, \alpha A
x^{2-\delta})$,  we obtain for gluon PDF:
\begin{eqnarray} \z g(x, Q^2)  = -
\frac{{( \xi_{sg})}^{\delta}}{ r^{1+\delta}_{sg}}  \biggl[ 2
\cdot
\frac{d F_2(x \xi_{sg}, Q^2)}{dlnQ^2}
+ c_1F_2(x \xi_{sg},Q^2)  + c_2F_2(Ax \xi_{sg},Q^2) \nonumber \\ \z
{}~~~~~~ ~~~~~
+~  O(x^{2-\delta}, \alpha A x^{2-\delta}) \biggr]
\label{6}\end{eqnarray}

Because the value of $A$ is arbitrary it is convenient to neglect the
contribution from $F_2(x \xi_{sg}/A,Q^2)$. Putting
formally $A=\infty$\footnote{Really $A \sim  (x \xi_{sg})^{-1}$.}, one
arrives to
the final general formula to extract $g(x,Q^2)$ in the NLO approximation:
\begin{eqnarray} g(x, Q^2)  = -
\frac{{( \xi_{sg})}^{\delta}}{ r^{1+\delta}_{sg}}  \biggl[ 2
\cdot
\frac{d F_2(x \xi_{sg}, Q^2)}{dlnQ^2}
{}~+~ r^{1+\delta}_{ss}~F_2(x \xi_{sg},Q^2) ~+~
O(x^{2-\delta}, \alpha  x^{1-\delta}) \biggr]
\label{7}\end{eqnarray}

Restricting the analysis to $O(x^{2-\delta}, \alpha
 x^{1-\delta})$, one can replace
$ \xi_{sg} \to  \xi =
\gamma_{sg}^{(0),1+\delta}/\gamma_{sg}^{(0),\delta} $ into Eq. (\ref{7}):
\begin{eqnarray} g(x, Q^2)  = -
\frac{ \xi ^{\delta}}{ r^{1+\delta}_{sg}}  \biggl[ 2 \cdot
\frac{d F_2(x \xi, Q^2)}{dlnQ^2}
{}~+~ r^{1+\delta}_{ss}~F_2(x \xi,Q^2) ~+~
O(x^{2-\delta},\alpha x^{1-\delta}) \biggr]
\label{8}\end{eqnarray}

This replacement is very useful. The NLO AD
$\gamma_{sp}^{(1),n}$ are
singular\footnote{In the case of replacement Mellin convolution
  by ordinary product these singularities transform to logarithmically
  increasing terms (see \cite{KOTIYF93} and \cite{KOTIPRD94})}
 in both points, $n=1$ and $n=0$, and their presence
into the arguments of $\tilde p(x,Q^2)$ makes the numerical
agreement between this approximate formula and the exact calculation worse
(we have
checked this point using some MRS sets of parton
distributions).

Using NLO approximation of $r^{1+\delta}_{sp}$ we easily obtain the
final results for $g(x,Q^2)$:
\begin{eqnarray} \z  g(x, Q^2)  = - \frac{2f}{\alpha e}
\frac{ \xi ^{\delta}}{ \gamma^{(0),1+\delta}_{sg} +
\overline \gamma^{(1),1+\delta}_{sg} \alpha }  \biggl[
\frac{d F_2(x \xi, Q^2)}{dlnQ^2} \nonumber \\ \z
{}~~~~~~~~~~ ~+~ \frac{\alpha }{2} \cdot
\left\{ \begin{array}{lll}
 \Bigl( \gamma^{(0),1+\delta}_{ss} +
\overline \gamma^{(1),1+\delta}_{ss} \alpha  \Bigr) F_2(x \xi,Q^2) &+&
O(x^{2-\delta},\alpha x^{1-\delta}) \\
 \gamma^{(0),1+\delta}_{ss}
F_2(x \xi,Q^2) &+&
O(\alpha^2,x^{2-\delta},\alpha x^{1-\delta})
\end{array} \right.
\biggr]
\label{9.1} \\
\z  g(x, Q^2)  = - \frac{2f}{\alpha e}
\frac{ 1}{ \gamma^{(0),1+\delta}_{sg} +
\overline \gamma^{(1),1+\delta}_{sg} \alpha }  \biggl[
\frac{d F_2(x, Q^2)}{dlnQ^2} \nonumber \\ \z
{}~~~~~~~~~~ ~+~ \frac{\alpha }{2} \cdot
\left\{ \begin{array}{lll}
 \Bigl( \gamma^{(0),1+\delta}_{ss} +
\overline \gamma^{(1),1+\delta}_{ss} \alpha  \Bigr) F_2(x,Q^2) &~+&
O( x^{1-\delta}) \\
 \gamma^{(0),1+\delta}_{ss}
F_2(x,Q^2) &~+&
O(\alpha^2, x^{1-\delta})
\end{array} \right.
\biggr]
\label{9.2}
\end{eqnarray}
where
$$\overline \gamma^{(1),\eta}_{sg} ~=~  \gamma^{(1),\eta}_{sg} +
B_2^{g,\eta} \bigl(2\beta_0 +
  \gamma^{(0),\eta}_{gg} - \gamma^{(0),\eta}_{ss} \bigr) \biggr)$$
$$\overline  \gamma^{(1),\eta}_{ss} ~=~
 \gamma^{(1),\eta}_{ss} + B_2^{g,\eta} \gamma^{(0),\eta}_{gs} +
  2\beta_0 B_2^{s,\eta}$$

 Any equation from above formulae (10) may be used, because there is
 a strong cancelation between the shifts in the arguments of the function
 $F_2$ and its derivative and the shifts in the coefficients in
front of them.

For concrete values of $\delta = 0.5$ (see also \cite{KOPA})
and $\delta = 0.3$ we obtain (for f=4 and $\overline{MS}$ scheme):
\begin{eqnarray} \z
{}~\mbox{ if }~ \delta =0.5 \nonumber \\ \z
g(x, Q^2)  = \frac{105}{4 e \sqrt{23 \times 77}} \frac{1}{\alpha}
\frac{1}{ (1 + 26.93 \alpha) }  \biggl[
\frac{d F_2(\frac{23}{77} x , Q^2)}{dlnQ^2} \nonumber \\ \z
{}~~~~~~~~~~   ~+~ \frac{16}{3} \alpha \Bigl( \frac{107}{60} -2ln2 \Bigr)
\left\{ \begin{array}{lll}
\Bigl\{1+ 37.76 \alpha \Bigr\}
F_2(\frac{23}{77} x,Q^2) &+&
O(x^{2-\delta},\alpha x^{1-\delta}) \\
F_2(\frac{23}{77} x,Q^2)
 &+&
O(\alpha^2,x^{2-\delta},\alpha x^{1-\delta})
\end{array} \right.
\biggr]
\label{10.1} \\
\z  g(x, Q^2)  = \frac{105}{92 e } \frac{1}{\alpha}
\frac{1}{ (1 + 26.93 \alpha) }
  \biggl[
\frac{d F_2(x, Q^2)}{dlnQ^2} \nonumber \\ \z
{}~~~~~~~~~~ ~+~ \frac{16}{3} \alpha \Bigl( \frac{107}{60} -2ln2 \Bigr)
\left\{ \begin{array}{lll}
\Bigl\{1+ 37.76 \alpha \Bigr\}
F_2( x,Q^2)
 &~+&
O( x^{1-\delta}) \\
F_2(x,Q^2)
 &~+&
O(\alpha^2, x^{1-\delta})
\end{array} \right.
\biggr]
\label{10.2}
\\ \z \nonumber
\\ \z
\nonumber \\ \z
{}~\mbox{ if }~ \delta =0.3 \nonumber \\ \z
g(x, Q^2)  = \frac{0.60
}{\alpha e}
\frac{1}{ (1 + 52.52 \alpha) }  \biggl[
\frac{d F_2( x/5.27 , Q^2)}{dlnQ^2} \nonumber \\ \z
{}~~~~~~~~~~   ~+~ 1.89 \alpha
\left\{ \begin{array}{lll}
\Bigl\{1- 50.67 \alpha \Bigr\}
F_2( x/5.27,Q^2) &+&
O(x^{2-\delta},\alpha x^{1-\delta}) \\
F_2(x/5.27,Q^2)
 &+&
O(\alpha^2,x^{2-\delta},\alpha x^{1-\delta})
\end{array} \right.
\biggr]
\label{10.3} \\
\z  g(x, Q^2)  = \frac{0.98
}{\alpha e}
\frac{1}{ (1 + 52.52 \alpha) }
  \biggl[
\frac{d F_2(x, Q^2)}{dlnQ^2} \nonumber \\ \z
{}~~~~~~~~~~ ~+~ 1.89 \alpha
\left\{ \begin{array}{lll}
\Bigl\{1- 50.67 \alpha \Bigr\}
F_2( x,Q^2)
 &~+&
O( x^{1-\delta}) \\
F_2(x,Q^2)
 &~+&
O(\alpha^2, x^{1-\delta})
\end{array} \right.
\biggr]
\label{10.4}
\end{eqnarray}\\

{\bf 2.}
 Assuming the {\it non-Regge-like behaviour} for the gluon distribution and
 $F_2(x,Q^2)$:
$$g(x,Q^2)  =  \frac{
\exp{(\frac{1}{2} \sqrt{\delta_g(Q^2)ln\frac{1}{x}})}  }{
{(2\pi \delta_g(Q^2)ln\frac{1}{x})}^{1/4}       }
 \tilde g(x,Q^2),~~
F_2(x,Q^2)  =  \frac{
\exp{(\frac{1}{2} \sqrt{\delta_s(Q^2)ln\frac{1}{x}})}  }{
{(2\pi \delta_s(Q^2)ln\frac{1}{x})}^{1/4}       }
 \tilde f(x,Q^2), $$
 we obtain the
following equation for the $Q^2$ derivative of
the SF $F_2$\footnote{Using a lower approximation $O(x)$ is not very exact,
because in this case $F_2$ and the gluon distribution can contain an
additional factor in the form of a serie $1+ \sum_k
(1/\delta_p/ln(1/x))^k$,
which is determined by boundary conditions (see discussion in
Ref.\cite{BF}). We will not consider the appearance of this factor in
our analysis}:
 \begin{eqnarray} \frac{dF_2(x,Q^2)}{dlnQ^2}  =
-\frac{1}{2}
 \sum_{p=s,g} \frac{
\exp{(\frac{1}{2} \sqrt{\delta_p(Q^2)ln\frac{1}{x}})}  }{
{(2\pi \delta_p(Q^2)ln\frac{1}{x})}^{1/4}       }
\Bigl(  \tilde
 r^{1}_{sp}(\alpha) ~\tilde p(0,Q^2) +
 O(x^{1}) \Bigr) ,\label{13} \end{eqnarray}
 where $ \tilde r^{1 }_{sp}(\alpha) $ can be obtained from
 corresponding functions
$ r^{1+\delta }_{sp}(\alpha) $ replacing the singular term $1/\delta$
at $\delta \to 0$ by another term $1/\tilde \delta$:
\begin{eqnarray}
\frac{1}{\delta} \stackrel{\delta \to 0}{\to} \frac{1}{\tilde \delta}
{}~=~ \sqrt{\frac{ln(1/x)}{\delta_p(Q^2)}}
- \frac{1}{4\delta_p(Q^2)} \left[ 1 + \sum_{m=1}^{\infty}
\frac{1 \times 3 \times ... \times
(2m-1)}{\left(4\sqrt{\delta_p(Q^2) \ln(1/x)}\right)^m} \right]
\label{14} \end{eqnarray}
 The singular term appears only in the NLO part of the AD
$\gamma^{(1),1+\delta}_{sp}$ in Eq. (\ref{2.2}).
The replacement (\ref{14}) corresponds to the following transformation:
\begin{eqnarray}
\gamma^{(1),1+\delta}_{sp} \equiv \hat
\gamma^{(1),1}_{sp} \frac{1}{ \delta} +\breve \gamma^{(1),1+\delta}_{sp}
{}~~ \stackrel{\delta \to 0}{\to} ~~  \tilde \gamma^{(1),1}_{sp} = \hat
\gamma^{(1),1}_{sp} \frac{1}{\tilde \delta} +\breve \gamma^{(1),1}_{sp},
 \label{15} \end{eqnarray}
where $\hat \gamma^{(1),1}_{sp}$ and $\breve \gamma^{(1),1+\delta}_{sp}$
are the coefficients corresponding to singular and regular parts of
$\gamma^{(1),1+\delta}_{sp}$, respectively.

We restrict here our calculations to $O(x)$ because at
$O(x^2)$ one obtains an additional factor:
$$\frac{
\exp{(\frac{1}{2} \sqrt{\delta_g(Q^2)ln\frac{1}{x \xi_{sg}}})}  }{
{(2\pi \delta_g(Q^2)ln\frac{1}{x \xi_{sg}})}^{1/4}       }
\frac{
{(2\pi \delta_g(Q^2)ln\frac{1}{x})}^{1/4}       }{
\exp{(\frac{1}{2} \sqrt{\delta_g(Q^2)ln\frac{1}{x}})}  }$$
in front of the function $F_2$ and its derivative. This factor
complicates very much the final formulae.

Repeating the analysis of the
previous section step by step using the replacement (\ref{15}), we get
(for f=4):
\begin{eqnarray} \z
  g(x, Q^2)  = \frac{3}{4 e} \frac{1}{\alpha}
\frac{1}{ (1 + 26 \alpha [
1/\tilde \delta - \frac{41}{13}])  }
  \biggl[
\frac{d F_2(x, Q^2)}{dlnQ^2} \nonumber \\ \z
{}~~~~~~~~~~ ~+~
\left\{ \begin{array}{lll}
 \alpha^2 \Bigl\{ 203
- 61/\tilde \delta
\Bigl\}
F_2( x,Q^2)
 &~+&
O( x^{1}) \\
0 &~+&
O(\alpha^2, x^{1})
\end{array} \right.
\biggr],
\label{11.1}
\end{eqnarray}
because
$\gamma^{(0)1}_{ss}=0$.\\

{\bf 3.}
In the case $x^{-\delta} \gg Const$ our formula (\ref{9.2}) at
accuracy $O(x^{1-\delta})$ coincides with the corresponding
Ellis-Kunszt-Levin result
from ref. \cite{EKL}. By other part, for $\delta =1$ one arrives to
\begin{eqnarray} \z
g(x, Q^2)  = \frac{3}{4 e
\alpha}
\frac{1}{ (1 + \frac{619}{108} \alpha) }  \biggl[
\frac{d F_2(x/2 , Q^2)}{dlnQ^2} \nonumber \\ \z
{}~~~~~~~~~~   ~+~ \frac{32}{9} \alpha
\left\{ \begin{array}{lll}
\Bigl\{1+ \Bigl[ \frac{26231}{1728}- \frac{1181}{576}f \Bigr] \alpha \Bigr\}
F_2(x/2,Q^2) &+&
O(x^{2-\delta},\alpha x^{1-\delta}) \\
F_2( x/2,Q^2)
 &+&
O(\alpha^2,x^{2-\delta},\alpha x^{1-\delta})
\end{array} \right.
\biggr],
\label{12.1}
\end{eqnarray}
that coincides with Prytz results \cite{PRYTZ} in the LO approximation, when
we neglect the contributions $\sim F_2(x,Q^2)$. Both formulae,
((\ref{12.1}) and one from \cite{PRYTZ}), are
similar to ours in the NLO case, too. Certainly, the value $\delta =1$ lies
outside the more standard predicted range $0 \leq \delta \leq 1/2$, however
in the case of large $\delta $ ($\delta
> 0.25$) the
final formula (\ref{8}) depends very slowly from the concrete value of
$\delta$. This is due to the strong cancelation between the shifts
in the arguments  and in the coefficients in front of the functions.

Equations (\ref{9.2}) and (\ref{11.1}) can be combined in a more general
formula valid for any value of $\delta$:
\begin{eqnarray}
\z  g(x, Q^2)  = - \frac{2f}{\alpha e}
\frac{ 1}{ \gamma^{(0),1+\delta}_{sg} +
\tilde \gamma^{(1),1+\delta}_{sg} \alpha }  \biggl[
\frac{d F_2(x, Q^2)}{dlnQ^2} \nonumber \\ \z
{}~~~~~~~~~~ ~+~ \frac{\alpha }{2} \cdot
\left\{ \begin{array}{lll}
 \Bigl( \gamma^{(0),1+\delta}_{ss} +
\tilde \gamma^{(1),1+\delta}_{ss} \alpha  \Bigr) F_2(x,Q^2) &~+&
O( x^{1-\delta}) \\
 \gamma^{(0),1+\delta}_{ss}
F_2(x,Q^2) &~+&
O(\alpha^2, x^{1-\delta})
\end{array} \right.
\biggr]
\label{13.1}
\end{eqnarray}
where
$\tilde \gamma^{(1),1+ \delta}_{sg}$ coincides with $\overline
\gamma^{(1),1+\delta}_{sg}$  with the replacement:
\begin{eqnarray}
\frac{1}{\delta} \to \int^1_x \frac{dy}{y}~ \frac{g(y,Q^2)}{g(x,Q^2)}
\label{14.1}
\end{eqnarray}

In the cases $x^{-\delta} \gg Const$ and $\delta \to 0$ the r.h.s. of
(\ref{14.1}) leads to $1/\delta $ and $1/\tilde \delta $,
respectively.\\

{\bf 4.}
In Fig. 1 it is shown the accuracy of Eqs. (\ref{10.1}),
(\ref{10.3}) (both O($\alpha^2,x^{2-\delta},\alpha x^{1-\delta}$))
and (\ref{11.1}) (O($\alpha^2, x^{1}$)) in
the reconstruction of various gluon distributions from MRS sets
at Q$^2$=20 GeV$^2$.
We have chosen for this test MRS(D$_{0}$) ($\delta$=0), MRS(D$_{-}$)
($\delta$=0.5) and MRS(G) ($\delta$=0.3) as three representative
densities (see ref. \cite{MRS95} and references therein).
It can be observed in Fig. 1a that using the formula with $\delta=0.5$
one gets the best
agreement with the input parameterization (less than 1 $\%$) in the
case of MRS(D$_{-}$) set;
for MRS(G) the
reconstruction is still good (less than 10 $\%$), but for
MRS(D$_{0}$) the deviation reach a 30 $\%$ at low x.

In Fig. 1b the degree of accuracy of the reconstruction formula
with $\delta=0.3$ can be observed. Here one should expect
the set MRS (G) to give
also a very good ($\sim 1\%$ level)
agreement,
however this is not the case because set (G)
distinguishes the exponents of the sea-quark part $\delta_s \sim 0$
from the gluon density ($\delta=0.3$). Thus, Eq. (12) might be slightly
modified to treat this case.

Figs. 1c and 1d deal with the case $\delta=0$. As in Fig. 1a, one
can observe a very good accuracy in the reconstruction when $Q_0^2$
coincides with that of the test parameterization (4 $GeV^2$ for MRS set).
Notice also the lost of accuracy at high $x$ due
to the importance of the O($x$) terms neglected in Eq. (16)

With the help of Eq. (\ref{10.2}) we have extracted the gluon distribution
from HERA data, using the slopes dF$_2$/dlnQ$^2$
determined in ref. \cite{H1PREP95} and ref. \cite{ZEUSGLU95}.
When H1 data are used the value of F$_2$ in Eq. (\ref{10.2}) was
directly taken from the parameterization given by H1 in ref. \cite{F2H1}.
With ZEUS data we substitute directly the F$_2$ values
presented in table 1 of their ref. \cite{ZEUSGLU95}.
We have checked that the
use of the H1 parameterization for $F_2$ when dealing with ZEUS data,
does not change significantly the $xG(x,Q^2)$ result.

Figures  2a and 2b shows the extracted values of the gluon distribution.
It can be observed that the agreement
within the errors between the bands, generated from a global fit to data,
the parameterization MRS(G), and the extracted points is excellent.\\

{\bf 5.}
In conclusion, a set of new formulae connecting the gluon density
with F$_2$ at
low x have been presented. They work fairly well for singular type gluons
($\delta \sim 0.5-0.3$) and for the non-singular case ($ \delta \sim 0$).
We have reproduced previous results of Prytz \cite{PRYTZ} using $\delta =1$,
and of Ellis-Kunszt-Levin \cite{EKL} with accuracy $O(x^{1-\delta})$.

We have found that for singular type of gluons the results do not depend
practically on the concrete value of the slope $\delta $:
there is a cancelation between the changes in the arguments and in
coefficients in front of the functions. However, when
$\delta \to 0$ the coefficients in front of
$dF_2(x,Q^2)/dlnQ^2$ and $F_2(x,Q^2)$ have singularities
leading to terms $\sim \sqrt{ln(1/x)}$\footnote{This happens
in the framework of
the double-logarithmic asymptotic. The singularities lead to terms
$\sim ln(1/x)$ in the case of Regge-like asymptotic (see
\cite{KOTILOWX}, \cite{KOTIYF93}, \cite{KOTIPRD94}).}. Consequently,
before to apply these formulae, some fit (maybe quite crude) of
experimental data is necessary to verify the type of $F_2(x)$
asymptotic at $x \to 0$.

The formulae were used to generate the gluon distribution
that agree with the rise observed by H1 and ZEUS experiments.
Further work is in progress in
order to
obtain similar
expressions connecting F$_L$, F$_2$ and  the Q$^2$ derivative of F$_2$.

\vskip 0.5 cm
%
%
%
%
\noindent{\bf Acknowledgments }

This work was supported in part by CICYT.
We are grateful to J.W. Stirling for providing
the parton distributions used in this work, and to
P. Aurenche and J. Kwiecinski for discussions.
%
%

%
%
%
\noindent{\bf Figure captions }
\vspace{1.cm}

\noindent{Figure 1:
Relative difference between the reconstructed gluon distribution
using formulae in text and different input parameterizations}

\vspace{1.cm}

\noindent{Figure 2: The gluon density. The points were extracted
from Eq. (\ref{10.2}) using H1 (Fig. 2a) and ZEUS (Fig. 2b) data.
The dashed curves shows the limits of
the error band taken from Fig. 3a of
paper \cite{H1PREP95} and Fig. 4 in ref. \cite{ZEUSGLU95}
which represents the uncertainty from a NLO fit. Solid line is the gluon
density from set MRS(G) \cite{MRS95}}

\vspace{1.cm}
\end{document}